\documentclass[aps,english,twocolumn,pre,floatfix]{revtex4-2}
\usepackage[T1]{fontenc}
\usepackage{amsmath}
\usepackage{amssymb}
\usepackage{lineno,hyperref}
\usepackage[latin9]{inputenc}
\usepackage{graphicx}
\usepackage{esint}
\usepackage{color}
\usepackage{lipsum}
\usepackage{accents}
\usepackage{xcolor}
\usepackage[normalem]{ulem}
\usepackage{appendix}
\usepackage{etoolbox}
\apptocmd{\thebibliography}{\raggedright}{}{}
\newcommand{\dint}{\displaystyle \int}

\makeatletter

\usepackage{babel}

\makeatother

\begin{document}

\title{A generalized Haus master equation model for mode-locked class-B lasers}

\author{Michel Nizette$^{a}$}
\email{michel.nizette@gmail.com}

\author{Andrei G. Vladimirov$^{b}$}
 \email{vladimir@wias-berlin.de}

\affiliation{$^{a}$D{\'e}partement de Physique, Facult{\'e} des Sciences, Universit{\'e} Libre de Bruxelles,\ CP 231, Campus Plaine, B-1050 Bruxelles, Belgium}
\affiliation{$^{b}$Weierstrass Institute, Mohrenstr. 39, 10117 Berlin, Germany}
\date{\today}

\begin{abstract}
Using an asymptotic technique we develop a generalized version of class-B Haus partial differential equation mode-locking model that accounts for both the slow gain response to the averaged value of the field intensity and the fast gain dynamics on the scale comparable to the pulse duration. We show that unlike the conventional class-B Haus mode-locked  model, our model is able to describe not only Q-switched instability of the fundamental mode-locked regime, but also the leading edge instability leading to harmonic mode-locked regimes with the increase of the pump power.
\end{abstract}
\maketitle

%


\section{Introduction\label{SecIntro}}
The first successful generation of a stable continuous train of passively mode-locked laser pulses was reported almost fifty years ago, in a dye laser with a saturable dye \cite{Ippen72a}. From then on, passive mode-locking has been recognized as a very powerful technique for generating high-quality picosecond and sub-picosecond pulses with fast repetition rates, and ongoing efforts are still devoted to the design of improved devices producing shorter and shorter pulses.

Owing to the early theoretical analyses of New \cite{New74a} and Haus \cite{Haus75a,Haus75b}, the essential physical process responsible for passive mode-locking is well understood.  If the absorber saturates faster than the gain medium on arrival of a pulse, a short temporal window of positive net gain is opened, enabling the pulse amplification necessary to compensate for round-trip losses. The net gain then quickly becomes negative again as a consequence of the saturation of the gain medium, combined to the relaxation of the absorber back to it its unsaturated state if the latter is fast enough. This process favors emission in the form of narrow pulses. 

This basic physical picture has been confirmed by numerical and analytical studies of passively mode-locked laser models of various complexities \cite{ippen1994principles,Dubbeldam97a,akhmediev1998ultrashort,Avrutin00a,
paschotta2001passive,Vladimirov04a,Vladimirov05a,rossetti2011modeling,
grelu2012dissipative,wang2013comparison,marconi2014lasing,javaloyes2016cavity}. From the early work cited above \cite{Haus75a,Haus75b} originated a highly successful, universally adopted model for pulse amplification and shaping that is now known as Haus master equation \cite{Haus00a}.
It is a partial differential equation (PDE) that describes the temporal evolution of the pulse profile with successive round-trips. Coupled to appropriate rate equations for the gain and absorber dynamics, it provides a model that is able to reproduce and quantify the pulse amplification scenario described above. Due to its mathematical simplicity, Haus master equation has been tremendously useful in understanding passive mode-locking.

The same mechanisms that are at play in passive mode-locking are also responsible for a detrimental, but practically unavoidable physical effect known as Q-switching instability  \cite{Haus76a,Kaertner95a,Hoenniger99a,schibli2000suppression,
Rachinskii06a,Kolokolnikov06a,rafailov2013ultrafast,kudelin2020pulse}
of a mode-locked regime. This process is the hallmark of class-B lasers, where the photon lifetime in the cavity is comparable to or shorter than the gain recovery time, and characterized by the build-up of a modulation of the laser response with a period typically extending over many round-trips.  
Understanding the origin of Q-switching instability in a mode-locked laser and determining the conditions under which it can be avoided is of primary importance for the successful generation of regular trains of mode-locked pulses. Theoretical studies based on Haus master equation have been very useful to this end \cite{Kaertner95a,kartner1996soliton,kartner1998mode,Hoenniger99a,Kolokolnikov06a}.
However, the ability of the model to describe Q-switching requires some particular care in the formulation of the rate equation for the gain. The successful formulation is one where the gain recovery is assumed much slower than the cavity round-trip time (class-B laser) and responds only to an average value of laser intensity over time.

Unfortunately, this conventional Q-switching-enabled variant of the Haus model
is unable to describe the fast gain depletion-recovery cycle that accompanies successive pulse amplifications. Therefore, it cannot account for the gain contribution to pulse shaping, which is a serious shortcoming of the model. The situation is not too bad if the absorber is fast enough to follow the pulse intensity profile adiabatically, i.e. in the so-called fast-absorber case. This occurs, for example, when the saturable absorption is provided via additive-pulse mode-locking \cite{Ippen89a} or Kerr lensing \cite{Spence91a}. Then, the absorber saturates during pulse amplification and recover immediately afterwards, which suffices to create the short window of net gain needed for mode-locking; the contribution from the gain medium is not essential to the process.

However, the shortcoming mentioned above is critical if the absorber recovers on a time scale much longer than the pulse duration (the slow-absorber case). This situation occurs, for example, in a laser with dye \cite{Arthurs88a} or semiconductor \cite{Dubbeldam97a} absorber. In that case, the absorber remains saturated for a while after the passing of a pulse so that, in the absence of a significant depletion of the gain medium, the net gain would remain positive for a time interval significantly longer than the pulse duration. This scenario is incompatible with the simple idea that the net gain remains positive for only a short time, thereby favoring narrow pulses. It is also certainly questionable from a physical point of view, as a positive net gain outside of the temporal boundaries of the pulse, and the amplification of the residual field that would result, is generally thought to favor a detrimental build-up of a macroscopic field, leading to the eventual destruction of the mode-locking pattern. This phenomenon is known as background instability \cite{Vladimirov05a}.

The problem has been partly identified in \cite{Dubbeldam97a}, where it has been stated that 
absorber saturation alone cannot explain the mode-locking of a semiconductor laser. The full absorber saturation-and-recovery cycle, they argued, should be taken into account in the description of the amplification mechanism, implying that a semiconductor laser cannot be analyzed under the usual slow-absorber approximation 
neglecting the slow recovery processes on a time scale comparable to the pulse duration. Their non-approximate treatment of the problem indeed solved the issue at hand and led to a physically consistent picture of the pulse amplification cycle. However, their theory still lacks the ability to account for a fast response of the gain medium, which we regard as a limitation of the model rather than a physical reality. Numerical studies based on delay differential equation (DDE) model 
indeed suggest that the gain medium does respond on the pulse time scale, even in class-B lasers \cite{Vladimirov05a}. 


The purpose of this work is to overcome the shortcoming of the the conventional class-B laser Haus model  \cite{Kolokolnikov06a}, which does not reveal any gain dynamics on the pulse time scale. Using an approach based on multiscale expansion we derive a model for the gain dynamics that generalizes the conventional formulations and which, coupled to Haus master equation, describes both Q-switching and the gain depletion-recovery cycle in a satisfactory way. The benefit is  that a single version of Haus model allows the study of both phenomena.  Unlike the empirical extended Haus mode-locking model of Ref. \cite{Hausen20} our model is derived using a multiscale approach and the gain evolution on the slow time scale is included as an additional equation, rather then boundary condition. Furthermore, one can note that the discussion in \cite{Hausen20} is incomplete since it does not consider the conventional class-B models of soliton \cite{lederer1999multipulse} and passive
\cite{Haus75a,Kolokolnikov06a} mode-locking, which include slow gain evolution equations. Similarly to the coherent Haus model derived in \cite{perego2020coherent} our model contains two separate equations for the  slow and fast gain components. However, our model derived rigorously using the multiscale method is simpler than that reported in \cite{perego2020coherent}.

 We make our objective more precise in Sec. \ref{SecModels} by introducing the two conventional versions of the Haus model discussed above and the underlying assumptions, as well as examining their respective limitations from a more mathematical point of view. We define the classifications into class-A and class-B lasers and into fast and slow absorbers in terms of relative magnitudes of model parameters, and state the corresponding usual approximations. We pinpoint one of the assumptions of the model as the source of its shortcomings. We then derive, in Sec. \ref{SecAnal}, a new model for the gain dynamics with the limiting assumption relaxed, which is our main result. We only give the outline of the method; the calculations themselves are too lengthy to be included in the main body of this paper and are relegated to Appendices. In Sec. \ref{SecValid}, we check the validity of the new model formulation with the help of numerical simulations, and emphasize its ability to predict both Q-switching and the appearance of harmonic mode-locking regimes. Conclusions are given in Sec. \ref{SecConcl}.

\section{Conventional model formulations\label{SecModels}}

In order to understand how two different versions of Haus partial-differential equation model arise for different types of lasers and why both fall short of including all the relevant physics, it is helpful to review
briefly their derivations from more fundamental principles (see also Appendix A of \cite{Kolokolnikov06a} for a more detailed treatment in the particular case of class-B lasers). Our starting point is a difference-differential model for passive mode-locking due to Haus
\cite{Dubbeldam97a}: 
\begin{subequations}
\label{hausDDE} 
\begin{align}
a\left(T+r\right)-a\left(T\right) & =\frac{1}{2}\left(d^{2}\frac{\mathrm{d}^{2}}{\mathrm{d}T^{2}}+g-q-k\right)a\left(T\right)\label{hausDDE_a}\\
\frac{\mathrm{d}g}{\mathrm{d}T} & =\gamma_{g}\left(g_{\max}-g\right)-s_{g}ga^{2},\label{hausDDE_b}\\
\frac{\mathrm{d}q}{\mathrm{d}T} & =\gamma_{q}\left(q_{\max}-q\right)-s_{q}qa^{2},\label{hausDDE_c}
\end{align}
\end{subequations}
where $T$ represents time, $a$ denotes the instantaneous amplitude of the laser field, and $g$ and $q$ stand respectively for the gain and saturable losses per round-trip. The parameters $g_{\max}$ and $q_{\max}$ denote their unsaturated values. The parameter $r$ represents the cold cavity round-trip time (that is, the round-trip time for a weak fluctuation of the field inside the cavity at transparency) and $d$ is responsible for the spectral filtering due to the finite bandwidth of the optical cavity. 
$k$ represents the linear cavity losses, so that the net gain per round-trip is given by $g-q-k$. Finally, $\gamma_{g}$ and $\gamma_{q}$ are the relaxation rates of the gain and absorber media, and $s_{g}$ and $s_{q}$ are saturation coefficients.  Note that that although the difference-differential model (\ref{hausDDE}) 
is free from the limitations of the PDE Haus models discussed below it also has an important drawback: due to the presence of the second derivative in the RHS of Eq.~(\ref{hausDDE_a}) the smoothness of its solution is reduced each round trip. Physically this means that the high frequency perturbations of the solution grow with the round trip number.  

\subsection{Field equation}

Haus partial-differential master equation is easily derived as a limit of the difference-differential field equation (\ref{hausDDE_a}) for a large cavity bandwidth and a weak net gain. To this end, we introduce
a formal smallness parameter $\varepsilon$, in terms of which we define the scales of the gain and absorber variables $g$ and $q$
and of the pulse duration measure $d$ by performing the following substitutions in the model (\ref{hausDDE}): 
\begin{subequations}
\label{hausScaling} 
\begin{gather}
g\rightarrow\varepsilon^{2}g,\quad g_{\max}\rightarrow\varepsilon^{2}g_{\max},\label{hausScaling_a}\\
q\rightarrow\varepsilon^{2}q,\quad q_{\max}\rightarrow\varepsilon^{2}q_{\max},\label{hausScaling_b}\\
k\rightarrow\varepsilon^{2}k,\label{hausScaling_c}\\
d\rightarrow\varepsilon d.\label{hausScaling_d}
\end{gather}
\end{subequations}
disappear from the final equations. 
We further introduce a two-scale expansion for the time variable: 
\begin{equation}
\frac{\mathrm{d}}{\mathrm{d}T}=\frac{\partial}{\partial t}+\varepsilon^{2}\frac{\partial}{\partial\tau},\label{hausTimeExp}
\end{equation}
in terms of which the advanced variable 
in the field equation (\ref{hausDDE_a}) can be expressed as
\begin{eqnarray}
a\left(T+r\right)=a\left(t+r,\tau+\varepsilon^{2}r\right)\nonumber\\
\simeq a\left(t+r,\tau\right)+\varepsilon^{2}r\frac{\partial a}{\partial\tau}\left(t+r,\tau\right).\label{hausDelayExp}
\end{eqnarray}
Substituting the scaling (\ref{hausScaling}) and then the expansions
(\ref{hausTimeExp}) and (\ref{hausDelayExp}) into the difference-differential
equation (\ref{hausDDE_a}) and keeping corrections only to the order
of $\varepsilon^{2}$ gives
\begin{eqnarray}
a\left(t+r\right)-a\left(t\right)+\varepsilon^{2}r\frac{\partial a}{\partial\tau}\left(t+r\right)\nonumber\\
=\frac{1}{2}\varepsilon^{2}\left(d^{2}\frac{\partial^{2}}{\partial t^{2}}+g-q-k\right)a\left(t\right).\label{hausDDEExp}
\end{eqnarray}
Finally, equating separately the coefficients of like powers of $\varepsilon$
on either side of Eq. (\ref{hausDDEExp}) gives the following two
equations:
\begin{equation}
r\frac{\partial a}{\partial\tau}=\frac{1}{2}\left(d^{2}\frac{\partial^{2}}{\partial t^{2}}+g-q-k\right)a,\quad a\left(t+r\right)=a\left(t\right).\label{haus}
\end{equation}
The first one is the Haus partial-differential master equation, and
the second one provides a periodic boundary condition for it.

Note how Eq. (\ref{haus}) involves the two different time scales
$t$ and $\tau$ as independent variables. The validity of this two-dimensional representation of time relies on the property of quasi-continuous pulse evolution between successive round-trips. This means that the temporal profile of the pulse inside the cavity varies little from one round-trip to the next, as a consequence of the weak net gain assumption. The difference between the amplitudes of two successive emitted copies of the pulse in the left-hand side of Eq. (\ref{hausDDE_a})
then appears in Eq. (\ref{haus}) approximated as a continuous derivative,
where $\tau$ thus represents the slow time variable in terms of which the pulse evolution and shaping processes are described (or, more
generally, any process that takes place over several round-trips). In contrast, the role of the fast time variable $t$ is to express the instantaneous configuration of the field in the cavity at a particular stage of its evolution, as well as any other process that occurs on the time scale of the round-trip or faster. The periodic boundary
condition in Eq. (\ref{haus}) reflects the approximate periodicity of the pulse train over a few round-trips. Eq. (\ref{haus}) is the
simplest possible formulation of Haus master equation. Extensions exist that account for the complex nature of the field amplitude $a$
(to include phase dynamics) and other physical effects (such as group velocity dispersion, Kerr effect, or linewidth enhancement factors)
\cite{Haus00a}. However the specific problem addressed in this paper does not require such extensions (nor does it preclude their use).

\subsection{Absorber equation}

The gain $g$ and absorber $q$ must be described by their own evolution equations in order to provide a closed dynamical system together with Haus master equation (\ref{haus}). The absorber equation is obtained
trivially by substituting Eqs. (\ref{hausScaling}) and (\ref{hausTimeExp}) into Eq. (\ref{hausDDE_c}) and then retaining only the leading-order
terms in $\varepsilon$, which gives an equation that is formally identical to Eq. (\ref{hausDDE_c}), but with $T$ replaced with $t$:
Furthermore since this equation 
is linear in $q$, its 
general solution is the sum of a periodic contribution with the same period as $a^{2}$ and an exponentially decaying term that lasts only for a few round-trips.
With the restriction to solutions in which the transient contribution has already died out, we can thus impose periodic boundary condition on $q$. Therefore, we get: 
\begin{equation}
\frac{\partial q}{\partial t}=\gamma_{q}\left(q_{\max}-q\right)-s_{q}qa^{2},\quad q\left(t+r\right)=q\left(t\right).\label{abs}
\end{equation}

Two cases are usually distinguished about the time scales involved in Eq. (\ref{abs}). The first is that of a fast absorber, which refers to a situation where the absorber relaxes on a time scale much shorter
than the pulse duration: $\gamma_{q}^{-1}\ll d$. This time scale relationship allows the adiabatic elimination of the absorber variable
$q$ as an explicit function of the field intensity $a^{2}$, which is achieved by setting the left-hand side of the absorber rate equation
(\ref{abs}) to zero and solving it for $q$. This yields
\begin{equation}
q=\frac{q_{\max}}{1+\gamma_{q}^{-1}s_{q}a^{2}}.\label{absFast}
\end{equation}
The other distinguished case is that of a slow absorber, when the
absorber relaxes on a time scale much longer than the pulse duration,
but comparable to the round-trip time or shorter: $d\ll\gamma_{q}^{-1}\lesssim r$.
This situation often justifies the neglecting of the relaxation term
$\gamma_{q}\left(q_{\max}-q\right)$ in Eq. (\ref{abs}) during the
absorber depletion stage, so that the explicit solution for $q$ is
now a function of the cumulative field energy up to time $t$:
\begin{equation}
q=q_{1}\exp\left(-s_{q}\int_{t_{1}}^{t}\mathrm{d}t\ a^{2}\right),\label{absSlow}
\end{equation}
where $q_{1}$ represents the absorber state just before the pulse arrival, and $t_{1}$ is the corresponding instant in time. Eq. (\ref{absSlow})
holds for the duration of a pulse; after that, the neglected relaxation process takes over as the laser field vanishes. Note that for both
fast and slow absorbers, the relaxation process is assumed to occur on a time scale not longer than the round-trip time, and is consistently
described in Eq. (\ref{abs}) in terms of the fast time variable $t$ rather than the slow time variable $\tau$.

\subsection{Gain equation}

For the gain medium, a similar dichotomic classification based on the relaxation rate $\gamma_{g}$ exists, but the reference time scale is different. A laser for which the gain relaxation takes place on a time scale comparable to the round-trip time or slower ($\gamma_{g}^{-1}\lesssim r$) is called a class-A laser. In contrast, a laser whose gain medium relaxes over many round-trips ($\gamma_{g}^{-1}\gg r$) is called
a class-B laser. This classification will be used to determine which time variable ($t$ or $\tau$) is involved in the description of the gain relaxation process. Unlike the absorber rate equation, we
shall see that there is no single formulation of the rate equation for the gain that will handle both cases, so one must choose from the outset which kind of laser is involved.

\subsubsection{Gain equation for a class-A laser}

In a class-A laser, the gain recovery time is not that long compared to the round-trip time. The derivation of the gain rate equation thus does not require any particular assumption on the relaxation and saturation rates $\gamma_{g}$ and $s_{g}$, and is entirely parallel
to that of the absorber rate equation (\ref{abs}).
We obtain:
\begin{equation}
\frac{\partial g}{\partial t}=\gamma_{g}\left(g_{\max}-g\right)-s_{g}ga^{2},\quad g\left(t+r\right)=g\left(t\right).\label{gainClassA}
\end{equation}
Together, Eqs. (\ref{haus}), (\ref{abs}), and (\ref{gainClassA})
form a closed system for the field, absorber, and gain medium in a
class-A laser.

The validity of Eq. (\ref{gainClassA}) does not extend to class-B lasers, however. A simple argument for this is that a correct class-B laser model should reduce, in the absence of fast mode-locking dynamics,
to the classical pair of rate equations that describes a single-mode emission \cite{Kaertner95a,Hoenniger99a,vladimirov2012modeling}.
 But while dropping the dependence in the fast time variable $t$ in Eq. (\ref{haus}) gives the correct rate equation  
for the field $a$, dropping it in Eq. (\ref{gainClassA}) gives a simple algebraic equation from which the gain can be solved as a function of the field intensity: 
\begin{equation}
g=\frac{g_{\max}}{1+\gamma_{g}^{-1}s_{g}a^{2}},\label{gainFast}
\end{equation}
instead of the expected rate equation. 
In fact, so far as $t$-independent solutions are considered, the explicit expression (\ref{absFast}) for $q$ holds no matter whether the absorber
is fast or slow, so Eqs. (\ref{absFast}) and (\ref{gainFast}) can be both substituted into the field rate equation 
to give a single closed rate equation for the field amplitude $a$ which cannot demonstrate oscillatory behavior.
This provides evidence (in the particular case of single-mode emission) that Q-switching cannot arise from the class-A formulation (\ref{gainClassA}) of the gain rate equation.

\subsubsection{Gain equation for a class-B laser}

A different equation for the gain is therefore required for a proper
description of Q-switching in a class-B laser. In order to account
for the slowness of the gain relaxation and saturation processes,
we must supplement the scaling (\ref{hausScaling}) with the following
substitution relations:
\begin{equation}
\gamma_{g}\rightarrow\varepsilon^{2}\gamma_{g},\quad s_{g}\rightarrow\varepsilon^{2}s_{g}.\label{hausScalingQSwitch}
\end{equation}
Substituting Eqs. (\ref{hausScaling}) and (\ref{hausScalingQSwitch})
into Eq. (\ref{hausDDE_b}) then gives
\begin{equation}
\frac{\mathrm{d}g}{\mathrm{d}T}=\varepsilon^{2}\left[\gamma_{g}\left(g_{\max}-g\right)-s_{g}ga^{2}\right].\label{hausGainEqExp1}
\end{equation}
The structure of Eq. (\ref{hausGainEqExp1}) justifies the application
of an averaging method \cite{Kevorkian96a}. This consists in taking
the gain variable $g$ as independent of the fast time variable $t$,
expressing the time derivative in the left-hand side in terms of the
slow time variable $\tau$ (using the relation $\tau=\varepsilon^{2}T$),
and averaging the right-hand side over one period in $t$. We thus
obtain:
\begin{equation}
\frac{\mathrm{d}g}{\mathrm{d}\tau}=\gamma_{g}\left(g_{\max}-g\right)-s_{g}gr^{-1}\int_{0}^{r}\mathrm{d}t\ a^{2}.\label{gainClassBInert}
\end{equation}

The form (\ref{gainClassBInert}) of the gain equation was used successfully in \cite{Kolokolnikov06a} to predict Q-switching. For particular solutions independent of the fast time $t$, Eq. (\ref{gainClassBInert}) reduces to the correct single-mode rate equation, 
so Eq. (\ref{gainClassBInert}) passes the simple validity test that the class-A gain equation (\ref{gainClassA}) did not pass. However, Eq. (\ref{gainClassBInert}) involves only the mean intensity over one round-trip, so according to it, the gain cannot respond to fast field variations. Although consistent with the assumption that the gain medium is much slower, this introduces a serious new limitation
in the model. Indeed, consider a mode-locked class-B laser with a slow absorber. In view of Eq. (\ref{absSlow}), the net gain $g-q-k$
during the passing of a pulse is given by
\begin{equation}
g-q-k=g-q_{1}\exp\left(-s_{q}\int_{t_{1}}^{t}\mathrm{d}t\ a^{2}\right)-k,\label{netGainMonotonous}
\end{equation}
where $g$ and $q_{1}$ are independent of $t$. This expression is a monotonously increasing function of $t$, 
consistently with the fact that the only dynamical process taken into account by Eq. (\ref{netGainMonotonous}) is the absorber saturation, which only contributes to a gradual increase of the net gain. As argued in the introduction, the monotonous net gain evolution is not confirmed (at least for common operating conditions)
by numerical simulations of the DDE mode-locked laser model, which suggest to the contrary that some fast dynamics of the gain medium does play a significant role in shaping the net gain profile, even in class-B lasers \cite{Vladimirov05a}.

\section{Improved model formulation\label{SecAnal}}

We have shown in the previous section that both the Q-switching-enabled gain model (\ref{gainClassBInert}) and the class-A gain model (\ref{gainClassA})
suffer shortcomings when applied to class-B lasers. While Eq. (\ref{gainClassA}) is unable to predict Q-switching, Eq. (\ref{gainClassBInert}) is unable to describe the fast response of the gain medium to the passing of a pulse. The two models in fact miss part of the physics for opposite reasons: in Eq.(\ref{gainClassBInert}), the gain medium is not fast enough to follow the fast intensity variations, whereas in Eq. (\ref{gainClassA}) it is not slow enough to endow the system with the necessary inertia to develop slow oscillations.

Since both Eqs. (\ref{gainClassA}) and (\ref{gainClassBInert}) are obtained as limits of the more general Eq. (\ref{hausDDE_b}), the key to obtaining a unified model capable of describing both phenomena
is to drop some of the scaling assumptions (\ref{hausScaling}) and (\ref{hausScalingQSwitch}).
To identify which ones can be retained and which are to be relaxed, we note that part of the success of the more complex model studied in \cite{Vladimirov05a} stems from its extended validity into the regime of strong amplification that typically holds in semiconductor
lasers. This observation suggests reconsidering the appropriateness of the weak-gain assumption (\ref{hausScaling_a}). A strong enough
pumping of the gain medium may indeed be required to compensate for its slow responsiveness to intensity variations and create a modulation
the gain profile of sufficient depth to induce a non-negligible contribution to pulse shaping. Moreover, both the pumping rate and the lasing threshold
usually influence the range of variation of the gain, which hints at the need to drop the weak cavity loss assumption (\ref{hausScaling_c})
too. In this section, 
basing on a set of assumptions weakened along those lines, we develop a generalized model for the gain dynamics with the desired properties.

Before proceeding, we note that the derivation of difference-differential model (\ref{hausDDE}) from fundamental principles already incorporates
an assumption of weak gain and losses \cite{Dubbeldam97a} (as is manifest from their linearity in $g$, $q$, and $k$). One may therefore
legitimately question its appropriateness as a starting point for an analysis that is intended to retain validity for a larger range
of gain than the classical theories presented in Sec. \ref{SecModels}.
To settle this point, we consider also the DDE model used in \cite{Vladimirov05a},
which holds for arbitrary gain and losses: 
\begin{subequations}
\label{andrei} 
\begin{gather}
\left(1+d\frac{\mathrm{d}}{\mathrm{d}T}\right)a\left(T+R\right)=K^{\frac{1}{2}}\exp\left[\frac{1}{2}\left(g-q\right)\right]a(T),\label{andrei_a}\\
\frac{\mathrm{d}g}{\mathrm{d}T}=\gamma_{g}\left(g_{\max}-g\right)-s_{g}\left[\exp\left(g\right)-1\right]\exp\left(-q\right)a^{2},\label{andrei_b}\\
\frac{\mathrm{d}q}{\mathrm{d}T}=\gamma_{q}\left(q_{\max}-q\right)-s_{q}\left[1-\exp\left(-q\right)\right]a^{2},\label{andrei_c}
\end{gather}
\end{subequations}
 where $a(T)$ is the electric field envelope at the entrance of the absorber medium, $g(T)$ and $q(T)$ are gain and loss introduced by the amplifying and absorber sections, respectively, and $T$ is time. $R$ is the cold cavity round trip time, $d$ is the inverse spectral filtering width, and $K$ is the attenuation factor per cavity round trip. The parameters $g_{max}$ and $q_{max}$ describe the unsaturated gain and absorption, while $s_{g}$ and $s_{q}$ are the saturation factors of the corresponding sections. 
For simplicity we have omitted in Eqs. (\ref{andrei}) the linewidth enhancement factors introduced in \cite{Vladimirov05a} to describe semiconductor lasers and assumed that $a$ is real. However, all the calculations below can be trivially generalized to the complex case when the linewidth enhancement factors are present.

In view of Eq. (\ref{andrei_a}), the quantity $g+\ln\left(K\right)$ represents the gain above linear cavity loss. Let us introduce new variable
$n$, such as 
\begin{equation}
n\left(1-\frac{n}{4}\right)=g+\ln\left(K\right).\label{GPrimeDef}
\end{equation}
Close to the threshold we can consider the following scaling 
\begin{subequations}
\label{scalingAndrei} 
\begin{gather}
n\rightarrow\varepsilon n,\label{scalingAndrei_a}\\
q\rightarrow\varepsilon^{2}q,\quad q_{\max}\rightarrow\varepsilon^{2}q_{\max},\label{scalingAndrei_b}\\
d\rightarrow\varepsilon d,\label{scalingAndrei_c}\\
\gamma_{g}\rightarrow\varepsilon^{2}\gamma_{g},\quad s_{g}\rightarrow\varepsilon^{2}s_{g}\label{scalingAndrei_d}
\end{gather}
\end{subequations}
 with small $\varepsilon$. Eq. (\ref{scalingAndrei_a}) can be viewed as a weakened form of the low-gain assumption (\ref{hausScaling_a}), as it allows the gain variations to cover a larger range (on the order of $\varepsilon$ instead of $\varepsilon^{2}$). Also, we allow arbitrarily large values of the pumping term $g_{\max}$ and of the linear losses $|\ln K|$. In all other respects, the scaling (\ref{scalingAndrei}) is identical to Eqs. (\ref{hausScaling}) and (\ref{hausScalingQSwitch}) together. Substituting the expression for $g$ obtained from Eq. (\ref{GPrimeDef}) together with the relation $R=r-\varepsilon d$ into Eqs. (\ref{andrei}), applying the scaling law (\ref{scalingAndrei}) and keeping only the lowest-order terms in $\varepsilon$ gives: 
\begin{subequations}
\label{andreiScaled1} 
\begin{gather}
a\left(T+r\right)-a\left(T\right)
=\frac{1}{2}\left(\varepsilon^{2}d^{2}\frac{\mathrm{d}^{2}}{\mathrm{d}T^{2}}+\varepsilon n-\varepsilon^{2}q\right)a(T)+O\left(\varepsilon^{3}\right),\label{andreiScaled1_a}\\
\frac{\mathrm{d}n}{\mathrm{d}T}=\varepsilon\left[P_{g}-\varepsilon\overline{\gamma}_{g}n-\left(S_{g}+\varepsilon\overline{s}_{g}n\right)a^{2}\right]+O\left(\varepsilon^{3}\right),\label{andreiScaled1_b}\\
\frac{\mathrm{d}q}{\mathrm{d}T}=\gamma_{q}\left(q_{\max}-q\right)-s_{q}qa^{2}+O\left(\varepsilon^{2}\right),\label{andreiScaled1_c}
\end{gather}
\end{subequations}
where we have used the relation $a\left(T+r\right)=a\left(T\right)+O\left(\varepsilon \right)$  and
\begin{eqnarray}
P_{g}=\gamma_{g}\left[g_{\max}+\ln\left(K\right)\right],\quad\overline{\gamma}_{g}=\gamma_{g}-\frac{1}{2}P_{g},\nonumber\\
 S_{g}=\left(K^{-1}-1\right)s_{g},\quad\overline{s}_{g}=\frac{1}{2}(3K^{-1}-1)s_{g}.\label{andreiNotations1}
\end{eqnarray}

 
Similarly, substituting $g=n+k$ into Eqs.~(\ref{hausDDE}) and using the scaling (\ref{scalingAndrei}) we get a system equivalent to Eqs.~(\ref{andreiScaled1}) up to $O\left(\varepsilon^{2}\right)$ corrections. This level of accuracy is sufficient to justify all calculations in this paper, which establishes the equivalence of the models (\ref{hausDDE}) and (\ref{andrei}) in the limit (\ref{scalingAndrei}).

The equations (\ref{andreiScaled1}) have been used to derive a generalized class B laser version of the Haus master equations. Note that since all quantities involved in the absorber equation (\ref{andreiScaled1_c}) are scaled as in  Eqs. (\ref{hausScaling}) and (\ref{hausScalingQSwitch}), the asymptotic absorber equation
(\ref{abs}) 
thus remain valid in the limit considered here, so Eq. (\ref{hausDDE_c}) does not require any further analysis. From now on, we focus all our efforts
on dealing with the remaining field equation (\ref{andreiScaled1_a}) and gain equation (\ref{andreiScaled1_b}). 
Because the calculations are too lengthy 
they are relegated to Supplemental Material where a
multiscale analysis is applied to Eqs. (\ref{andreiScaled1}) with small $\varepsilon$ in order to obtain the generalized version of the Haus master equations coupled to the gain rate equations.  
 Namely, we obtain
\begin{equation}
r\frac{\partial a}{\partial\tau}=\frac{1}{2}\left(d^{2}\frac{\partial^{2}}{\partial t^{2}}+n-q\right)a,\quad a\left(t+r\right)=a\left(t\right),\label{hausClassB}
\end{equation}
which are just Eq. (\ref{haus}) expressed in terms of the gain above
threshold $n$ together with Eqs. (\ref{abs}) 
and the system 
\begin{subequations}
\label{gainClassB}
\begin{align}
\frac{\partial n}{\partial t} & =S_{g}\left(r^{-1}\int_{0}^{r}a^{2}\mathrm{d}t-a^{2}\right),\label{gainClassB_a}\\
\frac{\mathrm{d}\overline{n}}{\mathrm{d}\tau} & =P_{g}-\overline{\gamma}_{g}\overline{n}-\left(S_{g}+\overline{s}_{g}\overline{n}\right)r^{-1}\int_{0}^{r}a^{2}\mathrm{d}t,\label{gainClassB_c}
\end{align}
\end{subequations}
 where $\overline{n}=r^{-1}\int_{0}^{r}n\mathrm{d}t$ represents the
average gain over one round-trip and the parameters $P_{g}$, $\overline{\gamma}_{g}$,
$S_{g}$, and $\overline{s}_{g}$ are defined by Es.~(\ref{andreiNotations1}). 

The rate equation (\ref{gainClassB_c}) is similar to the conventional gain rate equation (\ref{gainClassBInert}) for a class-B laser presented
in Sec. \ref{SecModels}, in that it involves the field intensity
averaged over one round-trip time, and is not sensitive to the details
of the mode-locked emission pattern. Those two equations would in
fact be completely equivalent, were it not for the presence of a different
saturation coefficient $\overline{s}_{g}$ in Eq. (\ref{gainClassB_c})
and for the fact that Eq. (\ref{gainClassB_c}) describes only the
mean value of the gain over one round-trip. Based on the knowledge
of the evolution of this mean value, Eq. (\ref{gainClassB_a}) determines
the full depletion-recovery cycle of the gain. The formulation (\ref{hausClassB})
of Haus master equation does not involve the $t$-independent solution
$\overline{n}$ of Eq. (\ref{gainClassB_c}) directly, but the $t$-dependent
solution $n$ of Eq. (\ref{gainClassB_a}) that averages to $\overline{n}$.
In that sense, Eqs. (\ref{gainClassB}) extend Eq. (\ref{gainClassBInert})
by accounting for the fast gain dynamics on a time scale comparable to the duration of a mode-locked pulse while retaining on average
the slow dynamics of Eq. (\ref{gainClassBInert}) responsible for Q-switching.

The correction (\ref{andreiNotations1}) to the saturation coefficient finds its justification in the multiple-scale expansion of the advanced
term $a(T+r)$ in Eq. (\ref{andreiScaled1_a}). According to the calculations of Appendix \ref{AppMultipleScale}, limiting the expansion to the
first derivative as in Eq. (\ref{hausDelayExp}) is not valid anymore
in the limit (\ref{scalingAndrei}). The second derivative does play
a role in the analysis, and leads to a contribution to the net gain
that is found to be equivalent to an effective decrease of the gain
saturability.

Substituting $n=g-k$ with $k=S_g/\overline{s}_g$ into Eqs.~(\ref{hausClassB}) and (\ref{gainClassB}), rescaling the field amplitude $a\to a/\sqrt{S_g}$ in the resulting equations, and combining them with the absorber equation (\ref{abs}) 
we get
\begin{subequations}
\label{ExtendedHaus}
\begin{gather}
r\frac{\partial a}{\partial\tau}=\frac{1}{2}\left(d^{2}\frac{\partial^{2}}{\partial t^{2}}+g-q-k\right)a,\quad a\left(t+r\right)=a\left(t\right),\label{ExtendedHaus1}\\
\frac{\partial q}{\partial t}=q_{0}-\gamma_{q}q-\overline{s}_{q}qa^{2},\quad q\left(t+r\right)=q\left(t\right),\label{ExtendedHaus2}\\
\frac{\partial g}{\partial t}=r^{-1}\int_{0}^{r}a^{2}\mathrm{d}t-a^{2},\label{ExtendedHaus3}\\
\frac{\mathrm{d}\overline{g}}{\mathrm{d}\tau}=g_{0}-\overline{\gamma}_{g}\overline{g}-\overline{g}(kr)^{-1}\int_{0}^{r}a^{2}\mathrm{d}t,\label{ExtendedHaus4}\\
\overline{g}=r^{-1}\int_{0}^{r}g\mathrm{d}t,
\label{ExtendedHaus5}
\end{gather}
\end{subequations}
where  $g_{0}=P_{g}+k\gamma_{g}$, $q_{0}=\gamma_{q}q_{\max}$, and $\overline{s}_{q}=s_{q}/S_{g}$.
Note that after substitution $g\to\overline{g}$  Eqs. (\ref{ExtendedHaus1}), (\ref{ExtendedHaus2}), and (\ref{ExtendedHaus4}) become formally equivalent to the conventional class-B Haus model.

\section{Numerical results\label{SecValid}}

An algorithm for solving numerically the extended Haus model that incorporates the gain equations (\ref{ExtendedHaus3}), (\ref{ExtendedHaus4}), and (\ref{ExtendedHaus5}) should proceed as follows. An initial condition for this problem is
the profile $a\left(t,\tau_{0}\right)$ of the field in the cavity
at $\tau=\tau_{0}$, which is periodic in $t$ with period $r$,
together with the single value $\overline{g}\left(\tau_{0}\right)$
of the mean gain at $\tau=\tau_{0}$. The absorber depletion-recovery
profile $q\left(t,\tau_{0}\right)$ is then determined from Eq. (\ref{ExtendedHaus2}) 
and the gain depletion-recovery profile $g\left(t,\tau_{0}\right)$ is computed from Eq. (\ref{ExtendedHaus3}). Since any solution of Eq. (\ref{ExtendedHaus3})
automatically has the same period as $a^{2}$, the periodicity of
$g\left(t,\tau_{0}\right)$ does not have to be imposed explicitly.
However, Eq. (\ref{ExtendedHaus3}) determines its solutions only up to an arbitrary additive constant, which is to be fixed by the
integral condition (\ref{ExtendedHaus5}).
The knowledge of $q\left(t,\tau_{0}\right)$ and $g\left(t,\tau_{0}\right)$
then provides enough data to compute the field profile $a\left(t,\tau_{0}+\mathrm{d}t\right)$ one time step $\mathrm{d}t$ later, using Haus master equation (\ref{ExtendedHaus1})
together with its periodic boundary condition. Likewise, the mean
gain $\overline{g}\left(\tau_{0}+\mathrm{d}t\right)$ one step later
is computed from Eq. (\ref{ExtendedHaus4}). Starting data is then
available for the next integration step. 
\begin{figure}[htp]
\centering \includegraphics[scale=0.4]{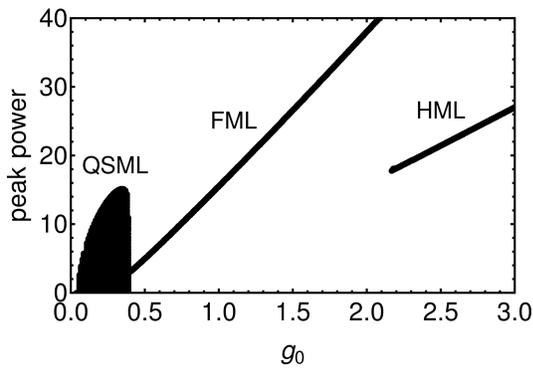} \caption{Pulse peak power as a function of the pump parameter $g_{0}$ obtained using the generalized Haus model, Eqs. (\ref{ExtendedHaus}). QSML, FML, and HML denote Q-switched, fundamental, and harmonic (with two pulses per cavity round trip) mode-locking regimes, respectively. Parameter values are: $r=2.5$, $k=0.519$, $q_{0}=1.0$, $\gamma_{g}=7.5\cdot10^{-3}$, $\gamma_{q}=0.2$, $s_{q}=7.0$, $d=0.02$.}
\label{fig:Bif}
\end{figure}
\begin{figure}[htp]
\centering \includegraphics[scale=0.4]{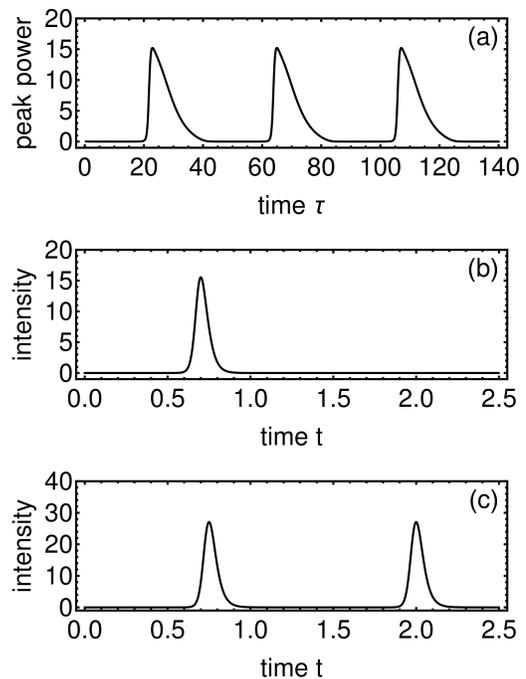} \caption{Time traces obtained by numerical integration of the generalized Haus model, Eqs. (\ref{ExtendedHaus}). (a) -- $g=0.3$, pulse peak power of the Q-switched mode-locking regime as function of slow time  $\tau$. (b) -- $g_0=1.0$, intensity $a^2$ of the fundamental mode-locked regime as function of the fast time $t$. (c) -- $g=3.0$, intensity $a^2$ of the harmonic mode-locking regime with two pulses per cavity round trip as a function of the fast time $t$.  Other parameters are the same as in Fig. \ref{fig:Bif}}
\label{fig:timetraces}
\end{figure}
\begin{figure}[htp]
\centering \includegraphics[scale=0.4]{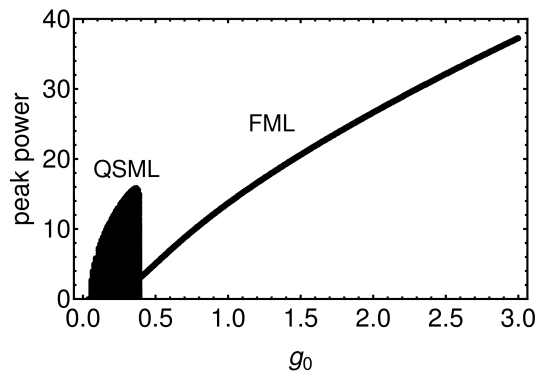} \caption{Bifurcation diagram similar to that shown in Fig. \ref{fig:Bif}, but obtained with conventional class-B Haus model. 
QSML and FML denote Q-switched and fundamental mode-locking regimes,
respectively. Parameter values are the same as in Fig. \ref{fig:Bif}.}
\label{fig:Bif-slow}
\end{figure}

\begin{figure}[htp]
\centering \includegraphics[scale=0.4]{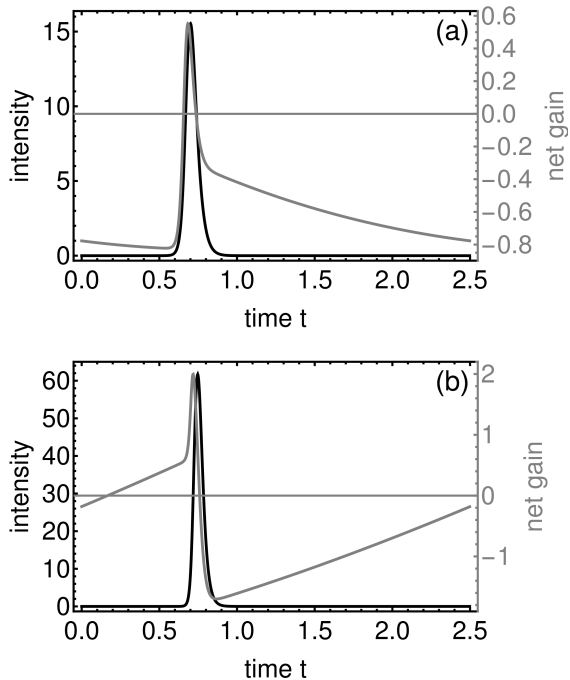} \caption{Pulse amplitude $a^2$ (black line) and net gain (gray line) obtained with the generalized Haus model (\ref{ExtendedHaus}) as functions of the fast time $t$. (a) -- $g_0=1.0$; (b) -- $g0=3.0$. Other parameters are the same as in Fig. \ref{fig:Bif}.}
\label{fig:net-gain}
\end{figure}
\begin{figure}[htp]
\centering \includegraphics[scale=0.4]{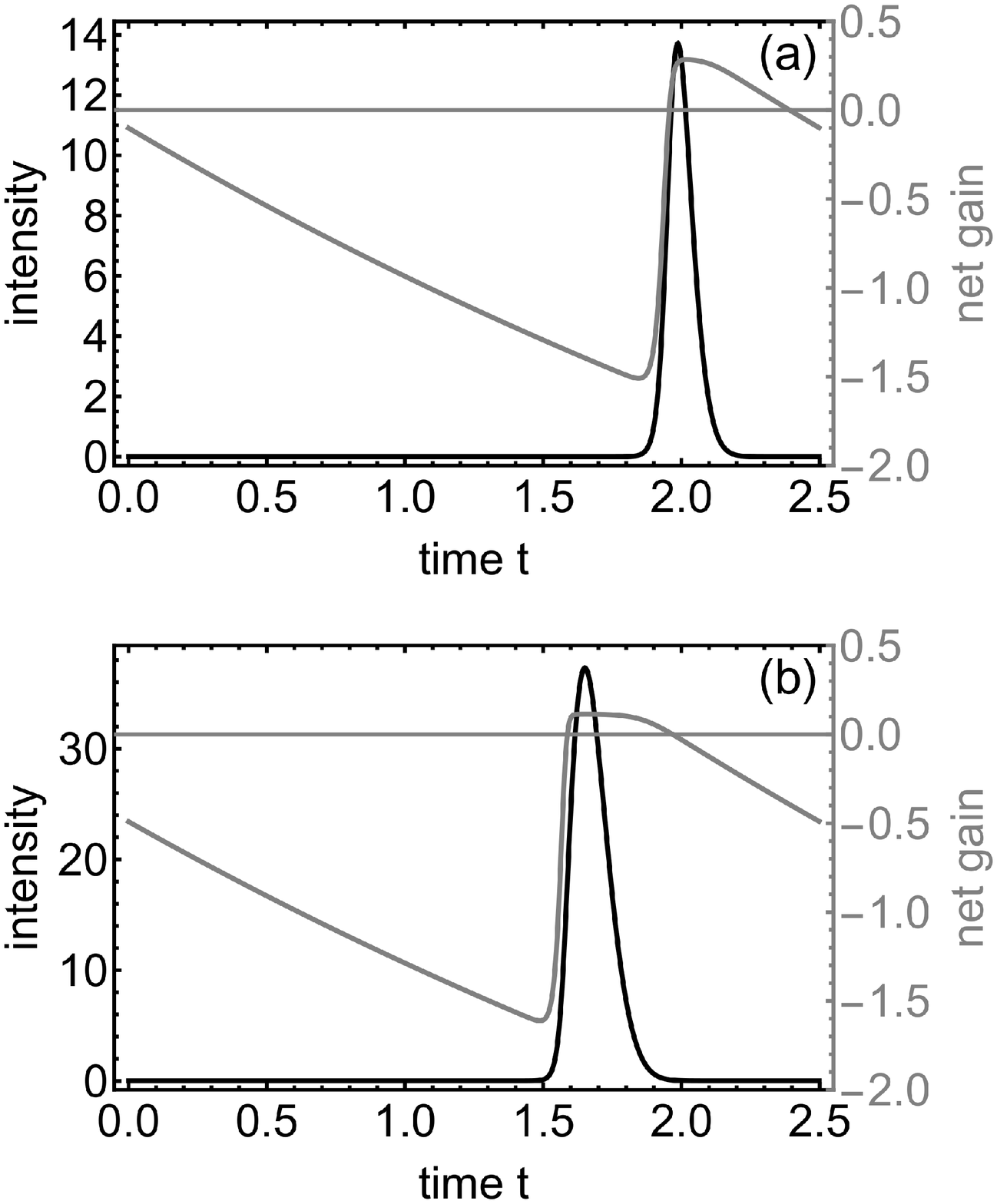} \caption{The same as Fig. \ref{fig:net-gain} but obtained with the conventional class-B Haus model.  (a) -- $g_0=1.0$; (b) -- $g0=3.0$. Other parameters are the same as in Fig. \ref{fig:Bif}.}
\label{fig:net-gain-slow}
\end{figure}
We have solved the generalized
Haus model (\ref{ExtendedHaus1})-(\ref{ExtendedHaus4})
numerically using the split-step method with $1024$ Fourier modes. The resulting bifurcation diagram presenting the evolution of pulse peak power $a^2$ with the increase of the pump parameter $g_{0}$ is shown
in Fig. \ref{fig:Bif}. It is seen that
apart from the fundamental mode-locked (FML) regime  with a single pulse per cavity round trip, this model can demonstrate harmonic mode-locking (HML) regime  with two pulses per cavity round trip time, as well as Q-switched mode-locking (QSML) regime  with periodically oscillating pulse peak power corresponding to a cloud of points in Fig. \ref{fig:Bif}. The slow time evolution of the pulse peak power of the QSML regime is shown in Fig. \ref{fig:timetraces}(a) together with the fast time evolution of the intensities of the FML and HML regimes, see Figs. \ref{fig:timetraces}(b) and \ref{fig:timetraces}(c), respectively.

Bifurcation diagram similar to that shown in Fig. \ref{fig:Bif}, but calculated using the conventional class-B Haus model,  which neglects the fast gain variation on the pulse width timescale is shown in Fig. \ref{fig:Bif-slow}. This model can be obtained by substituting $g\to\overline{g}$ into Eqs. (\ref{ExtendedHaus1}),  (\ref{ExtendedHaus2}), and (\ref{ExtendedHaus4}). 
It is seen that although the conventional model describes the Q-switched and fundamental mode-locking regimes rather well, it fails to describe the appearance of harmonic mode-locking regime, which emerges with the increase of the pump parameter $g_0$. Furthermore, the conventional model predicts slightly slower growth of the pulse peak power with $g_0$ and  broader mode-locked pulses than the generalized Haus model (\ref{ExtendedHaus}). 

Time dependence of the net gain  parameter $g-q-k$ [see Eq. (\ref{ExtendedHaus1})] and field intensity $a^2$ on
 the fast time $t$ is shown in Figs. \ref{fig:net-gain} and \ref{fig:net-gain-slow} for the generalized and
 conventional class-B Haus models, respectively. It is seen from Fig. \ref{fig:net-gain}(a) that in the generalized class-B Haus model with the pump parameter $g_0=1.0$  the net gain window corresponds to a short time interval when the pulse intensity is large. For larger pumps [see Fig.  \ref{fig:net-gain}(b) corresponding to $g_0=3.0$], however, positive net gain appears  before the pulse triggering the so-called {\it leading-edge  instability}  \cite{Vladimirov05a}, which eventually gives rise to a harmonic mode-locking regime. Since in the conventional class-B Haus model the gain is independent on the fast time $t$ the net gain parameter can only monotonously decrease between the pulses due to the absorber recovery. Therefore, this model cannot demonstrate the development of leading edge instability with the increase of the pump parameter, see Figs. \ref{fig:net-gain-slow}(a) and (b). Furthermore, since the fast gain saturation in absent in the conventional model, the net gain window is limited by the absorber recovery only. This is why the pulse widths obtained with the conventional model are broader than those of the generalized model (compare Figs. \ref{fig:net-gain-slow} and \ref{fig:net-gain}). Note that the physical mechanism of the development of the multipulse regimes reported in the soliton mode-locked lasers \cite{lederer1999multipulse} is different from the discussed above and requires the presence of dispersion and Kerr nonlinearity terms in the model equations.

\section{Conclusions\label{SecConcl}}
Although, unlike the DDE mode-locking model, the Haus master equations are based on low gain and loss approximation, which limits the parameter range of their validity, they are widely used and serve as an efficient tool for the analysis of mode-locked devices, such as e.g. fiber and solid state lasers. This is not only due to their simplicity and availability of well developed tools for analytical and numerical analysis of nonlinear PDEs, but also because of the possibility of straightforward inclusion of the group velocity dispersion into the master equations. On the contrary, the inclusion of the chromatic dispersion into the DDE mode-locking models is less straightforward, see \cite{pimenov2017dispersive,pimenov2020temporal}. Another limitation of the PDE Haus model, is that unlike difference-differential Haus equations (\ref{hausDDE}), the development of adequate PDE models of mode-locked class-B lasers require a careful formulation of the equations describing gain dynamics on different time scales. In particular, while the class A version of the Haus master equations (\ref{haus}), (\ref{abs}), and (\ref{gainClassA}) fails to describe 
Q-switching instability of the mode-locked regime, the conventional class-B model (\ref{haus}), (\ref{abs}), and (\ref{gainClassBInert}), which accounts for the slow time scale evolution of the gain, is capable of describing Q-switched mode-locking regime, but fails to predict the effect of gain on the pulse shaping as well as the leading edge pulse instability resulting in a transition to harmonic mode-locking regimes with the increase of the pumping parameter.  On the other hand, the solution of the difference-differential Haus model (\ref{hausDDE}), which is free from these  limitations, loses smoothness with increasing round trip number and hence exhibits an instability at large frequencies.  Here, using a rigorous asymptotic expansion technique, we have derived a generalized version of the Haus PDE model including the equations tor the slow and fast scale gain evolution which are simpler than similar equations reported in \cite{perego2020coherent}. By neglecting the gain evolution on the fast time scale our generalized equations can be transformed into the conventional class-B Haus model.  Our numerical simulations indicate that the generalized model allows to describe both the Q-switched mode-locking and the development the leading edge instability resulting in the appearance of a harmonic mode-locking regime.

\begin{acknowledgments}
The work of M. N. was supported by Fonds de la Recherche Scientifique--FNRS. The work of A. G. V.  was supported by the Deutsche Forschungsgemeinschaft   (DFG-RSF   project   No.445430311).
\end{acknowledgments}
\bibliographystyle{apsrev}

\pagebreak
\widetext
\numberwithin{equation}{section}
\renewcommand{\theequation}{S.\arabic{equation}}
\setcounter{page}{1}
\section{Supplemental Materials: A generalized Haus master equation model for mode-locked class-B lasers}

\subsection{\label{AppMultipleScale}Multiple-scale analysis of Eqs. (\ref{andreiScaled1})}

We first introduce a multiple-scale expansion for time. By analogy with Eq. (\ref{hausTimeExp}), we name the fastest time scale $t$ and the slower ones $\tau_{1}$ and $\tau_{2}$:
\begin{equation}
\frac{\mathrm{d}}{\mathrm{d}T}=\frac{\partial}{\partial t}+\varepsilon\frac{\partial}{\partial\tau_{1}}+\varepsilon^{2}\frac{\partial}{\partial\tau_{2}}+O\left(\varepsilon^{3}\right).\label{TExp}
\end{equation}
The advanced variable $a\left(T+r\right)$ in the field equation (\ref{andreiScaled1_a})
can be expressed in terms of the various time scales in the expansion
(\ref{TExp}) as follows:
\begin{eqnarray}
a\left(T+r\right)  =  a\left(t+r,\tau_{1}+\varepsilon r,\tau_{2}+\varepsilon^{2}r\right)=
\left[1+\varepsilon r\frac{\partial}{\partial\tau_{1}}+\varepsilon^{2}\left(\frac{r^{2}}{2}\frac{\partial^{2}}{\partial\tau_{1}^{2}}+r\frac{\partial}{\partial\tau_{2}}\right)\right]a\left(t+r,\tau_{1},\tau_{2}\right)
+O\left(\varepsilon^{3}\right).\label{delayExp}
\end{eqnarray}
We further expand: 
\begin{eqnarray}
a=a_{0}+\varepsilon a_{1}+\varepsilon^{2}a_{2}+O\left(\varepsilon^{3}\right),\quad n=n_{0}+\varepsilon n_{1}+\varepsilon^{2}n_{2}+O\left(\varepsilon^{3}\right).\label{varExp}
\end{eqnarray}
Substituting Eqs. (\ref{TExp}) and (\ref{delayExp}) and then Eq.
(\ref{varExp}) into Eqs. (\ref{andreiScaled1}) and equating the
coefficients of like powers of $\varepsilon$ separately leads to
a hierarchy of linear problems. A study of their solvability conditions
will provide a set of equations equivalent to Eqs. (\ref{andreiScaled1})
in the limit of small $\varepsilon$.

\subsubsection{\texorpdfstring{$O\left(\varepsilon^{0}\right)$}{Lg} problem}

The $O(\varepsilon^{0})$ problem is: 
\begin{subequations}
\label{order0} 
\begin{gather}
a_{0}\left(t+r\right)-a_{0}(t)=0,\label{order0_a}\\
\frac{\partial n_{0}}{\partial t}=0,\label{order0_b}
\end{gather}
\end{subequations}
 and gives a periodic boundary condition for the field amplitude $a_{0}$
and the information that the leading-order gain component $n_{0}$
does not vary on the fastest time scale $t$.

\subsubsection{\texorpdfstring{$O\left(\varepsilon^{1}\right)$}{Lg} problem}

The $O(\varepsilon^{1})$ problem is: 
\begin{subequations}
\label{order1} 
\begin{align}
a_{1}\left(t+r\right)-a_{1}(t) & =\left(-r\frac{\partial}{\partial\tau_{1}}+\frac{1}{2}n_{0}\right)a_{0},\label{order1_a}\\
\frac{\partial n_{1}}{\partial t} & =-\frac{\partial n_{0}}{\partial\tau_{1}}+P_{g}-S_{g}a_{0}^{2}.\label{order1_b}
\end{align}
\end{subequations}
 The validity of the expansions (\ref{varExp}) for all times requires
that $a_{1}$ and $n_{1}$ be bounded functions of $t$. This imposes
the vanishing of the right-hand side of Eq. (\ref{order1_a}) and
the vanishing of the average of the right-hand side of Eq. (\ref{order1_b})
over all $t$, which leads to the following solvability conditions:
\begin{subequations}
\label{solv1} 
\begin{gather}
r\frac{\partial a_{0}}{\partial\tau_{1}}=\frac{1}{2}n_{0}a_{0},\label{solv1_a}\\
\dfrac{\partial n_{0}}{\partial\tau_{1}}=P_{g}-S_{g}\widetilde{a}^{2},\label{solv1_b}
\end{gather}
\end{subequations}
 where $\widetilde{a}$ is the quadratic average field amplitude over
one round-trip: 
\begin{equation}
\widetilde{a}(\tau_{1},\tau_{2})=\sqrt{r^{-1}\int_{0}^{r}\mathrm{d}t\ a_{0}^{2}(t,\tau_{1},\tau_{2})}.\label{aTildeDef}
\end{equation}

The quantity $\widehat{a}$ defined by
\begin{equation}
a_{0}=\widetilde{a}\widehat{a}\label{a0Decomp}
\end{equation}
thus represents the emission pattern normalized so that
\begin{equation}
r^{-1}\dint_{0}^{r}\mathrm{d}t\ \widehat{a}^{2}=1.\label{a0Norm}
\end{equation}

It will be advantageous to express the field variable $a_{0}$ everywhere
in terms of the decomposition (\ref{a0Decomp}), as we will find $\widetilde{a}$
and $\widehat{a}$ to be governed separately by their own evolution
equations. First, the $O\left(\varepsilon^{0}\right)$ equation (\ref{order0_a})
for the field translates to a periodic boundary condition for $\widehat{a}$:
\begin{equation}
\widehat{a}\left(t+r\right)=\widehat{a}(t).\label{aHatPer}
\end{equation}

Next, substituting Eq. (\ref{a0Decomp}) into the solvability condition
(\ref{solv1_a}) for the field gives
\begin{equation}
r\left(\frac{\partial\widetilde{a}}{\partial\tau_{1}}\widehat{a}+\widetilde{a}\frac{\partial\widehat{a}}{\partial\tau_{1}}\right)=\frac{1}{2}n_{0}\widetilde{a}\widehat{a}.\label{solv12}
\end{equation}
Multiplying both sides of Eq. (\ref{solv12}) by $\widehat{a}$, integrating
over $t$, and using the normalization condition (\ref{a0Norm}) further
yields
\begin{equation}
r\dfrac{\partial\widetilde{a}}{\partial\tau_{1}}=\dfrac{1}{2}n_{0}\widetilde{a},\label{solv13}
\end{equation}
which, together with the solvability condition (\ref{solv1_b}) for
the gain, defines a closed system for the average field amplitude
$\widetilde{a}$ and leading-order gain $n_{0}$. This system is a
conservative oscillator that describes Q-switching and admits the
first integral
\begin{equation}
H\left(\tau_{2}\right)=\left(2r\right)^{-1}n_{0}^{2}+S_{g}\widetilde{a}^{2}-P_{g}\ln\left(P_{g}^{-1}S_{g}\widetilde{a}^{2}\right).\label{HDef}
\end{equation}
The slow evolution of the Q-switching energy $H$ on the slowest time
scale $\tau_{2}$ is as yet undetermined.

Substituting Eq. (\ref{solv13}) for $\widetilde{a}$ back into Eq.
(\ref{solv12}) gives
\begin{equation}
\frac{\partial\widehat{a}}{\partial\tau_{1}}=0,\label{aHatConst}
\end{equation}
which means that the normalized emission pattern $\widehat{a}$ does
not vary on the time scale $\tau_{1}$ of Q-switching. Finally, substituting
the solvability conditions (\ref{solv1}) and the decomposition (\ref{a0Decomp})
back into the $O\left(\varepsilon^{1}\right)$ problem (\ref{order1}),
we obtain 
\begin{subequations}
\label{order11} 
\begin{align}
a_{1}\left(t+r\right)-a_{1}(t)  =0,\label{order11_a}\\
\frac{\partial n_{1}}{\partial t}  =S_{g}\left(1-\widehat{a}^{2}\right)\widetilde{a}^{2}.\label{order11_b}
\end{align}
\end{subequations}
 Eq. (\ref{order11_a}) is a periodic boundary condition for $a_{1}$. Eq. (\ref{order11_b}) is an evolution equation for the small gain correction on the fast time scale $t$. Because its right-hand side averages to zero over one round-trip, its solutions $n_{1}$ are periodic
with period $r$.

Finally, by multiplying both sides of Eq.~(\ref{order11_b}) by $n_{1}$ and integrating over one period in $t$, we find an identity that will be useful later on: 
\begin{equation}
\int_{0}^{r}\mathrm{d}t\ n_{1}=\int_{0}^{r}\mathrm{d}t\ \widehat{a}^{2}n_{1}.\label{n1AveIdent}
\end{equation}

\subsubsection{\texorpdfstring{$O\left(\varepsilon^{2}\right)$}{Lg} problem}

The $O\left(\varepsilon^{2}\right)$ problem is: 
\begin{subequations}
\label{order2} 
\begin{align}
a_{2}\left(t+r\right)-a_{2}(t)  =\left(-r\frac{\partial}{\partial\tau_{1}}+\frac{1}{2}n_{0}\right)a_{1}
+\left[-\frac{r^{2}}{2}\frac{\partial^{2}}{\partial\tau_{1}^{2}}-r\frac{\partial}{\partial\tau_{2}}+\frac{1}{2}\left(d^{2}\frac{\partial^{2}}{\partial t^{2}}+n_{1}-q\right)\right]a_{0},\label{order2_a}\\
\frac{\partial n_{2}}{\partial t}  =-\frac{\partial n_{1}}{\partial\tau_{1}}-2S_{g}a_{0}a_{1}-\left(\frac{\partial}{\partial\tau_{2}}+\overline{\gamma}_{g}+s_{g}a_{0}^{2}\right)n_{0}.\label{order2_b}
\end{align}
\end{subequations}
 The following calculations involve many variable changes and make heavy use of the field amplitude decomposition (\ref{a0Decomp}), of the normalization condition (\ref{a0Norm}), of the Q-switching
oscillator equations (\ref{solv1_b}) and (\ref{solv13}), and of which variable is independent of which time scale. For the sake of concision, we omit from now on any reference to those in most places
where they are invoked.

Boundedness of $a_{2}$ and $n_{2}$ in $t$ requires the vanishing
of the right-hand side of Eq. (\ref{order2_a}) and of the average
of the right-hand side of Eq. (\ref{order2_b}) over all $t$, leading
to the solvability conditions 
\begin{subequations}
\label{solv2Pre} 
\begin{gather}
\left(r\frac{\partial}{\partial\tau_{1}}-\frac{1}{2}n_{0}\right)a_{1}-\frac{1}{2}n_{1}\widetilde{a}\widehat{a}
=\left[-\frac{r^{2}}{2}\frac{\partial^{2}}{\partial\tau_{1}^{2}}-r\frac{\partial}{\partial\tau_{2}}+\frac{1}{2}\left(d^{2}\frac{\partial^{2}}{\partial t^{2}}-q\right)\right]\widetilde{a}\widehat{a}\label{solv2Pre_a}\\
\frac{\partial}{\partial\tau_{1}}r^{-1}\int_{0}^{r}\mathrm{d}t\ n_{1}+2S_{g}\widetilde{a}r^{-1}\int_{0}^{r}\mathrm{d}t\ \widehat{a}a_{1}
=-\left(\frac{\partial}{\partial\tau_{2}}+\overline{\gamma}_{g}+s_{g}\widetilde{a}^{2}\right)n_{0}\label{solv2Pre_b}
\end{gather}
\end{subequations}
 Those can be rewritten as 
\begin{subequations}
\label{solv2} 
\begin{gather}
\left(r\frac{\partial a^{\prime}}{\partial\tau_{1}}-\frac{1}{2}n^{\prime}\right)\widehat{a} =\left[-r\widetilde{a}^{-1}\frac{\partial\widetilde{a}}{\partial\tau_{2}}+\frac{1}{2}\left(d^{2}\frac{\partial^{2}}{\partial t^{2}}-q\right)\right]\widehat{a},\label{solv2_a}\\
\frac{\partial\overline{n}^{\prime}}{\partial\tau_{1}}+2S_{g}\widetilde{a}^{2}\overline{a}^{\prime} =-\left(\frac{\partial}{\partial\tau_{2}}+\overline{\gamma}_{g}+\overline{s}_{g}\widetilde{a}^{2}+\frac{1}{2}n_{0}\frac{\partial}{\partial\tau_{1}}\right)n_{0},\label{solv2_b}
\end{gather}
\end{subequations}
 in terms of the effective saturation coefficient $\overline{s}_{g}$
given in Eq. (\ref{andreiNotations1}) and of the auxiliary variables
$a^{\prime}$, $\overline{a}^{\prime}$, $n^{\prime}$, and $\overline{n}^{\prime}$
defined as 
\begin{subequations}
\label{aux21} 
\begin{gather}
a^{\prime}=\widetilde{a}^{-1}\widehat{a}^{-1}a_{1},\quad\overline{a}^{\prime}=r^{-1}\int_{0}^{r}\mathrm{d}t\ \widehat{a}^{2}a^{\prime},\label{aux21_a}\\
n^{\prime}=n_{1}-\frac{r}{2}\left(P_{g}-S_{g}\widetilde{a}^{2}\right)-\frac{1}{4}n_{0}^{2},\quad\overline{n}^{\prime}=r^{-1}\int_{0}^{r}\mathrm{d}t\ n^{\prime}.\label{aux21_b}
\end{gather}
\end{subequations}
 Multiplying both sides of Eq.~(\ref{solv2_a}) by $\widehat{a}$,
integrating over one period in $t$, and using the identity (\ref{n1AveIdent})
yields, together with Eq.~(\ref{solv2_b}), the following linear
$t$-independent inhomogeneous system for two unknowns $\widetilde{a}^{\prime}$
and $\widetilde{n}^{\prime}$: 
\begin{subequations}
\label{solv2Bis} 
\begin{align}
r\frac{\partial\widetilde{a}^{\prime}}{\partial\tau_{1}}-\frac{1}{2}\widetilde{n}^{\prime} & =-r\widetilde{a}^{-1}\frac{\partial\widetilde{a}}{\partial\tau_{2}}-\frac{1}{2}r^{-1}\int_{0}^{r}\mathrm{d}t\ \widehat{a}^{2}q,\label{solv2Bis_a}\\
\frac{\partial\widetilde{n}^{\prime}}{\partial\tau_{1}}+2S_{g}\widetilde{a}^{2}\widetilde{a}^{\prime} & =-\left(\frac{\partial}{\partial\tau_{2}}+\overline{\gamma}_{g}+\overline{s}_{g}\widetilde{a}^{2}+\frac{1}{2}n_{0}\frac{\partial}{\partial\tau_{1}}\right)n_{0},\label{solv2Bis_b}
\end{align}
\end{subequations}
 where we have defined 
\begin{subequations}
\label{aux22} 
\begin{eqnarray}
\widetilde{a}^{\prime} & = & \overline{a}^{\prime},\label{aux22_a}\\
\widetilde{n}^{\prime} & = & \overline{n}^{\prime}-d^{2}r^{-1}\int_{0}^{r}\mathrm{d}t\ \left(\frac{\partial\widehat{a}}{\partial t}\right)^{2}.\label{aux22_b}
\end{eqnarray}
\end{subequations}
 Fredholm's solvability condition requires that the right-hand side
of the system (\ref{solv2Bis}) be orthogonal to the solutions of
the adjoint homogeneous problem. After a few calculations, we find
that this condition can be written in matrix form as 
\begin{gather}
\int_{0}^{T_{1}}\mathrm{d}\tau_{1}\left[\begin{array}{cc}
-2\left(P_{g}-S_{g}\widetilde{a}^{2}\right) & n_{0}\end{array}\right]
\times\left[\begin{array}{c}
-r\widetilde{a}^{-1}\dfrac{\partial\widetilde{a}}{\partial\tau_{2}}-\dfrac{1}{2}r^{-1}\dint_{0}^{r}\mathrm{d}t\ \widehat{a}^{2}q\\
-\left(\dfrac{\partial}{\partial\tau_{2}}+\overline{\gamma_{g}}+\overline{s}_{g}\widetilde{a}^{2}+\dfrac{1}{2}n_{0}\dfrac{\partial}{\partial\tau_{1}}\right)n_{0}
\end{array}\right]=0,\label{adjSol}
\end{gather}
where $T_{1}$ denotes the period of the Q-switching oscillator as
defined by Eqs. (\ref{solv1_b}) and (\ref{solv13}). Using the expression (\ref{HDef}) for the Q-switching energy $H$, Eq. (\ref{adjSol})
simplifies to
\begin{gather}
r\dfrac{\mathrm{d}H}{\mathrm{d}\tau_{2}}=T_{1}^{-1}\dint_{0}^{T_{1}}\mathrm{d}\tau_{1}\left[\left(P_{g}-S_{g}\widetilde{a}^{2}\right)r^{-1}\dint_{0}^{r}\mathrm{d}t\ \widehat{a}^{2}q\right.
-\left.\left(\overline{\gamma}_{g}+\overline{s}_{g}\widetilde{a}^{2}\right)n_{0}^{2}\right].\label{HEq}
\end{gather}
Eq. (\ref{HEq}) determines the slow evolution of the Q-switching
oscillation cycle. Its right-hand side involves the normalized emission
pattern $\widehat{a}$, whose evolution is as yet undetermined. We therefore now need an equation for $\widehat{a}$.

To this end, we multiply both sides of Eq.~(\ref{solv2Bis_a}) by
$\widehat{a}$ and subtract it side by side from Eq.~(\ref{solv2_a}),
obtaining
\begin{gather}
r\widehat{a}\frac{\partial\widehat{a}^{\prime}}{\partial\tau_{1}}=\left[-r\frac{\partial}{\partial\tau_{2}}\right.
\left.+\frac{1}{2}\left(d^{2}\frac{\partial^{2}}{\partial t^{2}}+\widehat{n}^{\prime}+r^{-1}\int_{0}^{r}\mathrm{d}t\ \widehat{a}^{2}q-q\right)\right]\widehat{a},\label{solv2Ter}
\end{gather}
where we have defined 
\begin{subequations}
\label{aux23} 
\begin{align}
\widehat{a}^{\prime} & =a^{\prime}-\widetilde{a}^{\prime},\label{aux23_a}\\
\widehat{n}^{\prime} & =n^{\prime}-\widetilde{n}^{\prime}.\label{aux23_b}
\end{align}
\end{subequations}
 Boundedness of $\widehat{a}^{\prime}$ in $\tau_{1}$ requires the
vanishing of the average of the right-hand side of Eq.~(\ref{solv2Ter})
over all $\tau_{1}$, leading to the following solvability condition:
\begin{gather}
r\dfrac{\partial\widehat{a}}{\partial\tau_{2}}=\dfrac{1}{2}\left[d^{2}\dfrac{\partial^{2}}{\partial t^{2}}+\widehat{n}\right.
\left.+T_{1}^{-1}\dint_{0}^{T_{1}}\mathrm{d}\tau_{1}\left(r^{-1}\dint_{0}^{r}\mathrm{d}t\ \widehat{a}^{2}q-q\right)\right]\widehat{a},\label{a0HatFast}
\end{gather}
where we have defined
\begin{equation}
\widehat{n}=T_{1}^{-1}\int_{0}^{T_{1}}\mathrm{d}\tau_{1}\widehat{n}^{\prime}.\label{aux24}
\end{equation}
Eq.~(\ref{a0HatFast}) provides an equation for $\widehat{a}$ that
depends on $\widehat{n}$.

A complementary equation for $\widehat{n}$ in terms of $\widehat{a}$
can be derived from the fast gain equation (\ref{order11_b}) using
the relations (\ref{aux21_b}), (\ref{aux22_b}), (\ref{aux23_b}),
and (\ref{aux24}) between the various auxiliary gain variables: 
\begin{equation}
\dfrac{\partial\widehat{n}}{\partial t}=S_{g}\left(1-\widehat{a}^{2}\right)T_{1}^{-1}\dint_{0}^{T_{1}}\mathrm{d}\tau_{1}\widetilde{a}^{2}.\label{n1HatFast}
\end{equation}

\subsection{Asymptotic form of Eqs. (\ref{andreiScaled1})\label{AppAsympt}}

We now summarize the results of the multiple-scale analysis of Appendix
\ref{AppMultipleScale} and collect all the obtained asymptotic equations
in a single place, renaming the leading-order gain component $n_{0}$
as
\begin{equation}
n_{0}=\widetilde{n}
\end{equation}
for the sake of notation uniformity. According to Eq. (\ref{a0Decomp}),
the field amplitude $a$ can be decomposed to leading order as the
product of a slow-varying envelope $\widetilde{a}$ and of a normalized
emission pattern $\widehat{a}$: 
\begin{equation}
a=\widetilde{a}\widehat{a},\label{asymptEqs_i}
\end{equation}
both governed by their own evolution equations.

Eqs. (\ref{a0HatFast}) and (\ref{n1HatFast}) form a system that
couples the emission profile $\widehat{a}$ to some fast gain component
$\widehat{n}$: 
\begin{subequations}
\label{asymptEqs_ab} 
\begin{align}
r\dfrac{\partial\widehat{a}}{\partial\tau_{2}}  =\dfrac{1}{2}\left[d^{2}\dfrac{\partial^{2}}{\partial t^{2}}+\widehat{n}\right.
\left.+T_{1}^{-1}\dint_{0}^{T_{1}}\mathrm{d}\tau_{1}\left(r^{-1}\dint_{0}^{r}\mathrm{d}t\ \widehat{a}^{2}q-q\right)\right]\widehat{a},\label{asymptEqs_a}\\
\dfrac{\partial\widehat{n}}{\partial t}  =S_{g}\left(1-\widehat{a}^{2}\right)T_{1}^{-1}\dint_{0}^{T_{1}}\mathrm{d}\tau_{1}\widetilde{a}^{2},\label{asymptEqs_b}
\end{align}
\end{subequations}
 where $T_{1}$ is the Q-switching period, to be defined more precisely below. Eq. (\ref{asymptEqs_a}) bears some similarity to Haus master equation (\ref{haus}), while Eq. (\ref{asymptEqs_b}) determines
the depletion-and-recovery profile of the gain over one round-trip. In view of Eqs. (\ref{aHatConst}) and (\ref{aux24}), $\widehat{a}$
and $\widehat{n}$ do not vary on the Q-switching time scale $\tau_{1}$, so they do not contain any information about Q-switching oscillations. In order to form a well-posed problem, the partial-differential system (\ref{asymptEqs_ab}) must satisfy some boundary conditions or other
constraints. Those are provided by Eqs. (\ref{a0Norm}) and (\ref{aHatPer}):
\begin{subequations}
\label{asymptEqs_cd} 
\begin{align}
\widehat{a}\left(t+r\right) & =\widehat{a}(t),\label{asymptEqs_c}\\
r^{-1}\dint_{0}^{r}\mathrm{d}t\ \widehat{a}^{2} & =1.\label{asymptEqs_d}
\end{align}
\end{subequations}
 Eq. (\ref{asymptEqs_c}) is a periodic boundary condition for the
emission pattern $\widehat{a}$. The solution $\widehat{n}$ of the fast gain equation (\ref{asymptEqs_b}) is defined up to an additive
contribution that depends only on the slowest time scale $\tau_{2}$, and whose value is to be adjusted so that the normalization condition
(\ref{asymptEqs_d}) holds for all $\tau_{2}$.

Coupled equations for the slow-varying field amplitude $\widetilde{a}$ and the leading-order gain component $\widetilde{n}$ are provided
by Eqs. (\ref{solv1_b}) and (\ref{solv13}): 
\begin{subequations}
\label{asymptEqs_ef} 
\begin{align}
r\dfrac{\partial\widetilde{a}}{\partial\tau_{1}} & =\dfrac{1}{2}\widetilde{n}\widetilde{a},\label{asymptEqs_e}\\
\dfrac{\partial\widetilde{n}}{\partial\tau_{1}} & =P_{g}-S_{g}\widetilde{a}^{2}.\label{asymptEqs_f}
\end{align}
\end{subequations}
 In view of Eqs. (\ref{order0_b}) and (\ref{aTildeDef}), $\widetilde{a}$
and $\widetilde{n}$ do not vary on the fast time scale $t$. Eqs.
(\ref{asymptEqs_ef}) define a conservative oscillator whose orbits
describe Q-switching cycles. Those are characterized by the values
$H$ of a first integral provided by Eq. (\ref{HDef}): 
\begin{equation}
H=\left(2r\right)^{-1}\widetilde{n}^{2}+S_{g}\widetilde{a}^{2}-P_{g}\ln\left(P_{g}^{-1}S_{g}\widetilde{a}^{2}\right).\label{asymptEqs_g}
\end{equation}
The Q-switching period $T_{1}$ can be computed as the period of $\widetilde{a}$
and $\widetilde{n}$ in $\tau_{1}$ according to Eqs. (\ref{asymptEqs_ef}).
The evolution of $\widetilde{a}$ and $\widetilde{n}$ on all time
scales, and thus the full Q-switching dynamics, is completely determined
by the additional knowledge of the evolution of the Q-switching energy
$H$ on the slowest time scale $\tau_{2}$. It is provided by Eq.
(\ref{HEq}): 
\begin{gather}
r\dfrac{\mathrm{d}H}{\mathrm{d}\tau_{2}}=T_{1}^{-1}\dint_{0}^{T_{1}}\mathrm{d}\tau_{1}\left[\left(P_{g}-S_{g}\widetilde{a}^{2}\right)r^{-1}\dint_{0}^{r}\mathrm{d}t\ \widehat{a}^{2}q\right.
\left.-\left(\overline{\gamma}_{g}+\overline{s}_{g}\widetilde{a}^{2}\right)\widetilde{n}^{2}\right],\label{asymptEqs_h}
\end{gather}
where $\overline{s}_{g}$ is given by Eq. (\ref{andreiNotations1}).

Finally, the system (\ref{asymptEqs_i})--(\ref{asymptEqs_h}) is
closed by its coupling to the absorber equation (\ref{abs}). 
The asymptotic equations (\ref{asymptEqs_i})--(\ref{asymptEqs_h}) can be useful in their own right as they would provide a good starting point for an all-analytical bifurcation study of mode-locked class-B lasers (though such a study falls outside of the scope of the present paper).
They are not the final result of the present analysis, however. Keeping in mind that our goal is to find a generalized gain model to be coupled to Haus
master equation (\ref{haus}), we note that Eqs. (\ref{asymptEqs_i})--(\ref{asymptEqs_h})
present the drawback of not involving the physical field and gain variables directly. The relations between those and the asymptotic dynamical variables are in fact rather complicated. As a second step
in the analysis, therefore, we perform various transformations (given in Appendix \ref{AppSumming}) to recast Eqs. (\ref{asymptEqs_i})--(\ref{asymptEqs_h})
into a much more physically transparent form. This procedure is not a strict application of asymptotic analysis as it involves inhomogeneous transformations (i.e., the summing of quantities proportional to distinct powers of $\varepsilon$). Nevertheless, it satisfies our goal by
yielding Haus master equation coupled to a new model for the gain
dynamics.

\subsection{\label{AppSumming}Equivalence of Eqs. (\ref{hausClassB})--(\ref{gainClassB})
to Eqs. (\ref{asymptEqs_i})--(\ref{asymptEqs_h})
in the limit (\ref{scalingAndrei})}

Eqs. (\ref{asymptEqs_i})--(\ref{asymptEqs_h}) involve the three independent time variables $t$, $\tau_{1}$, and $\tau_{2}$. Our
first step towards casting them into the form (\ref{hausClassB})--(\ref{gainClassB})
consists in recombining $\tau_{1}$ and $\tau_{2}$ into a single
slow-time variable $\tau$. To this end, our strategy is to propose
the following multiple-scale expansion for $\tau$:
\begin{equation}
\dfrac{\mathrm{d}}{\mathrm{d}\tau}=\frac{\partial}{\partial\tau_{1}}+\varepsilon\frac{\partial}{\partial\tau_{2}}+O\left(\varepsilon^{2}\right),\label{tauExp}
\end{equation}
and then devise (by means of educated guesswork) a set of equations
in the $t$ and $\tau$ variables that admit Eqs. (\ref{asymptEqs_i})--(\ref{asymptEqs_h})
as a limit for small $\varepsilon$. We then explicitly check the
correctness of that limit to establish formally the equivalence of
the two formulations. This can be viewed as a multiple scale analysis
applied backwards.

First, the equations (\ref{asymptEqs_ab}) for $\widehat{a}$ and
$\widehat{n}$ suggest 
\begin{subequations}
\label{hatEqs} 
\begin{align}
r\dfrac{\partial\widehat{a}}{\partial\tau} & =\dfrac{1}{2}\varepsilon\left[d^{2}\dfrac{\partial^{2}}{\partial t^{2}}+\widehat{n}^{\prime}+r^{-1}\dint_{0}^{r}\mathrm{d}t\ \widehat{a}^{2}q-q\right]\widehat{a},\label{hatEqs_a}\\
\dfrac{\partial\widehat{n}^{\prime}}{\partial t} & =S_{g}\left(1-\widehat{a}^{2}\right)\widetilde{a}^{2}\label{hatEqs_b}
\end{align}
\end{subequations}
 as equivalent for small $\varepsilon$. It is indeed easily checked
that the application of an averaging method to Eq. (\ref{hatEqs_a})
yields Eq. (\ref{asymptEqs_a}) with the definition (\ref{aux24})
for $\widehat{n}$ in terms of $\widehat{n}^{\prime}$, and that the
combination of Eqs. (\ref{aux24}) and (\ref{hatEqs_b}) leads to
Eq. (\ref{asymptEqs_b}).

Next, Eq. (\ref{asymptEqs_h}) for the slow evolution of Q-switching
oscillations suggest modifying the conservative oscillator (\ref{asymptEqs_ef})
as follows: 
\begin{subequations}
\label{tildeEqs} 
\begin{align}
r\dfrac{\partial\widetilde{a}}{\partial\tau} & =\dfrac{1}{2}\left(\widetilde{n}-\varepsilon r^{-1}\dint_{0}^{r}\mathrm{d}t\ \widehat{a}^{2}q\right)\widetilde{a},\label{tildeEqs_a}\\
\dfrac{\partial\widetilde{n}}{\partial\tau} & =P_{g}-S_{g}\widetilde{a}^{2}-\varepsilon\left(\overline{\gamma}_{g}+\overline{s}_{g}\widetilde{a}^{2}\right)\widetilde{n}.\label{tildeEqs_b}
\end{align}
\end{subequations}
 We now check that Eqs. (\ref{asymptEqs_ef})--(\ref{asymptEqs_h})
are a limit of Eqs. (\ref{tildeEqs}) for small $\varepsilon$ as
follows. We substitute the expansions (\ref{tauExp}) and 
\begin{equation}
\widetilde{a}=\widetilde{a}_{0}\left(1+\varepsilon\widetilde{a}^{\prime}\right)+O\left(\varepsilon^{2}\right),\quad\widetilde{n}=\widetilde{n}_{0}+\varepsilon\widetilde{n}^{\prime}+O\left(\varepsilon^{2}\right)\label{hatTildeExp}
\end{equation}
into Eqs. (\ref{tildeEqs}) and equate separately the coefficients
of like powers of $\varepsilon$, obtaining a hierarchy of problems
for the coefficients of the expansions (\ref{hatTildeExp}). The $O\left(\varepsilon^{0}\right)$
problem is equivalent to Eqs. (\ref{asymptEqs_ef}). The $O\left(\varepsilon^{1}\right)$
problem is 
\begin{subequations}
\label{tildeO1} 
\begin{align}
r\dfrac{\partial\widetilde{a}^{\prime}}{\partial\tau_{1}}-\frac{1}{2}\widetilde{n}^{\prime} & =-r\widetilde{a}_{0}^{-1}\dfrac{\partial\widetilde{a}_{0}}{\partial\tau_{2}}-\dfrac{1}{2}r^{-1}\dint_{0}^{r}\mathrm{d}t\ \widehat{a}^{2}q,\label{tildeO1_a}\\
\dfrac{\partial\widetilde{n}^{\prime}}{\partial\tau_{1}}+2S_{g}\widetilde{a}_{0}^{2}\widetilde{a}^{\prime} & =-\left(\dfrac{\partial}{\partial\tau_{2}}+\overline{\gamma}_{g}+\overline{s}_{g}\widetilde{a}^{2}\right)\widetilde{n}_{0},\label{tildeO1_b}
\end{align}
\end{subequations}
 and is formally identical to Eqs. (\ref{solv2Bis}) with the sole
exception of the absence of the last term in Eq. (\ref{solv2Bis_b}).
By the same reasoning as in Appendix \ref{AppMultipleScale}, therefore,
a solvability condition for Eqs. (\ref{tildeO1}) is provided by Eq.
(\ref{adjSol}) with the last term of the second element of the column
vector removed. Because that term vanishes in the integration over
one period in $\tau_{1}$, however, the presence of this term does
not matter, and the solvability condition simplifies to Eq. (\ref{HEq}).

The analysis so far establishes that in the limit for small $\varepsilon$,
the two time scales $\tau_{1}$ and $\tau_{2}$ can be recombined
into a single time variable $\tau$ by replacing the equations (\ref{asymptEqs_ab})
for $\widehat{a}$ and $\widehat{n}$ with Eqs. (\ref{hatEqs}) and
the equations (\ref{asymptEqs_ef})--(\ref{asymptEqs_h}) for $\widetilde{a}$
and $\widetilde{n}$ with Eqs. (\ref{tildeEqs}). We now want to cast
the two field equations (\ref{hatEqs_a}) and (\ref{tildeEqs_a})
into a single equation for the combined field amplitude $a$ given
by Eq. (\ref{asymptEqs_i}). To this end, we multiply both sides of
Eq. (\ref{hatEqs_a}) by $\widetilde{a}$ and both sides of Eq. (\ref{tildeEqs_a})
by $\widehat{a}$ and add the two resulting equations. Keeping in
mind that $\widetilde{a}$ does not depend on $t$, we obtain

\begin{equation}
r\dfrac{\partial a}{\partial\tau}=\dfrac{1}{2}\left(\varepsilon d^{2}\dfrac{\partial^{2}}{\partial t^{2}}+n-\varepsilon q\right)a,\label{hausClassBScaled}
\end{equation}
where we have defined
\begin{equation}
n=\widetilde{n}+\varepsilon\widehat{n}^{\prime}.\label{nSumming}
\end{equation}
Eq. (\ref{asymptEqs_c}) further yields a periodic boundary condition
for $a$:
\begin{equation}
a\left(t+r\right)=a\left(t\right).\label{hausClassBPerScaled}
\end{equation}

An equation for the recombined gain variable $n$ is then obtained
by differentiating both sides of Eq.\ (\ref{nSumming}) with respect
to $t$, substituting Eq. (\ref{hatEqs_b}), and keeping in mind that
$\widetilde{n}$ does not depend on $t$: 
\begin{equation}
\dfrac{\partial n}{\partial t}=\varepsilon S_{g}\left(r^{-1}\dint_{0}^{r}\mathrm{d}t\ a^{2}-a^{2}\right),\label{gainClassBScaled}
\end{equation}
where we have combined the field decomposition (\ref{asymptEqs_i})
and the normalization condition (\ref{asymptEqs_d}) to express $\widetilde{a}^{2}$
as follows:Append
\begin{equation}
\widetilde{a}^{2}=r^{-1}\dint_{0}^{r}\mathrm{d}t\ a^{2}.\label{aTildeRel}
\end{equation}

Eq. (\ref{gainClassBScaled}) alone determines $n$ only up to an
additive contribution that depends only on $\tau$. An extra constraint
is required to obtain a well-posed problem, and provided by differentiating
both sides of Eq.\ (\ref{nSumming}) with respect to $\tau$ and
integrating over one period in $t$, which gives
\begin{gather}
r^{-1}\dint_{0}^{r}\mathrm{d}t\ \dfrac{\partial n}{\partial\tau}=P_{g}-S_{g}\widetilde{a}^{2}
-\varepsilon\left(\overline{\gamma}_{g}+\overline{s}_{g}\widetilde{a}^{2}\right)r^{-1}\dint_{0}^{r}\mathrm{d}t\ \widetilde{n}+\varepsilon r^{-1}\dint_{0}^{r}\mathrm{d}t\ \dfrac{\partial\widehat{n}^{\prime}}{\partial\tau}\label{gainClassBAux1}
\end{gather}
Now, note that the equation (\ref{hatEqs_b}) for $\widehat{n}^{\prime}$
and the normalization condition (\ref{asymptEqs_d}) imply that $\widehat{n}^{\prime}$
is periodic in $t$ with the same period as $\widehat{a}^{2}$. A useful relation can be obtained by multiplying both sides of Eq. (\ref{hatEqs_a}) by $\widehat{a}$ and both sides of Eq. (\ref{hatEqs_b}) by $\widehat{n}^{\prime}$
and integrating the two resulting equations over one period in $t$.
This gives
\begin{equation}
\dint_{0}^{r}\mathrm{d}t\ \widehat{n}^{\prime}=\dint_{0}^{r}\mathrm{d}t\ \widehat{a}^{2}\widehat{n}^{\prime}=d^{2}\dint_{0}^{r}\mathrm{d}t\left(\dfrac{\partial\widehat{a}}{\partial t}\right)^{2},\label{gainRel1}
\end{equation}
which, in view of Eq. (\ref{hatEqs_a}), further entails
\begin{equation}
\dint_{0}^{r}\mathrm{d}t\ \frac{\partial\widehat{n}^{\prime}}{\partial\tau}=O\left(\varepsilon\right).\label{gainRel2}
\end{equation}
Substituting the expression (\ref{aTildeRel}) for $\widetilde{a}^{2}$,
the expression for $\widetilde{n}$ obtained from Eq. (\ref{nSumming}),
and Eq. (\ref{gainRel2}) into Eq. (\ref{gainClassBAux1}) gives 
\begin{gather}
\dfrac{\partial\overline{n}}{\partial\tau}=P_{g}-S_{g}r^{-1}\dint_{0}^{r}\mathrm{d}t\ a^{2}\nonumber\\
-\varepsilon\left(\overline{\gamma}_{g}+\overline{s}_{g}r^{-1}\dint_{0}^{r}\mathrm{d}t\ a^{2}\right)\overline{n}+O\left(\varepsilon^{2}\right),\label{gainAveClassBScaled}
\end{gather}
where $\overline{n}=r^{-1}\int_{0}^{r}n\mathrm{d}t$ represents the
average gain over one round-trip.

Finally, neglecting $O(\varepsilon^{2})$ terms in (\ref{gainClassBScaled})
and introducing the new variables $m=n+S_{g}/\overline{s}_{g}$ and
$\overline{m}=\overline{n}+S_{g}/\overline{s}_{g}$ and in we can
rewrite Eqs. (\ref{hausClassBScaled}), (\ref{hausClassBPerScaled}),
(\ref{gainClassBScaled}), and 
\begin{equation}
r\dfrac{\partial a}{\partial\tau}=\dfrac{1}{2}\left(\varepsilon d^{2}\dfrac{\partial^{2}}{\partial t^{2}}+m-\varepsilon q-k\right)a,\quad a\left(t+r\right)=a\left(t\right),\label{field}
\end{equation}
\begin{equation}
\dfrac{\partial m}{\partial t}=\varepsilon S_{g}\left(r^{-1}\dint_{0}^{r}\mathrm{d}t\ a^{2}-a^{2}\right),\label{gain}
\end{equation}
\begin{equation}
\dfrac{\partial\overline{m}}{\partial\tau}=P_{g}-S_{g}r^{-1}\dint_{0}^{r}\mathrm{d}t\ a^{2}-\varepsilon\left(\overline{\gamma}_{g}+\overline{s}_{g}r^{-1}\dint_{0}^{r}\mathrm{d}t\ a^{2}\right)\overline{n},\label{gainAveClassBScaled-1}
\end{equation}

Eqs. (\ref{hausClassBScaled}), (\ref{hausClassBPerScaled}), (\ref{gainClassBScaled}),
and (\ref{gainAveClassBScaled}) together are asymptotic to Eqs. (\ref{asymptEqs_i})--(\ref{asymptEqs_h})
in the limit for small $\varepsilon$. (The $O\left(\varepsilon^{2}\right)$
corrections in Eq. (\ref{gainAveClassBScaled}) can be safely neglected
without invalidating this result.) On the other hand, the same Eqs.
(\ref{hausClassBScaled}), (\ref{hausClassBPerScaled}), (\ref{gainClassBScaled}),
and (\ref{gainAveClassBScaled}) can be obtained by carrying out the
substitutions (\ref{scalingAndrei}) and $\tau\rightarrow\varepsilon^{-1}\tau$
into Eqs. (\ref{hausClassB})--(\ref{gainClassB}), which proves
the equivalence of Eqs. (\ref{hausClassB})--(\ref{gainClassB})
to Eqs. (\ref{asymptEqs_i})--(\ref{asymptEqs_h}) with the scaling
(\ref{scalingAndrei}).

\end{document}